\documentclass[journal]{IEEEtran}
\usepackage{cite}
\ifCLASSINFOpdf
  \usepackage[pdftex]{graphicx}
\else
\fi

\usepackage{xcolor}
\usepackage{array}
\begin{document}
%
% paper title
% Titles are generally capitalized except for words such as a, an, and, as,
% at, but, by, for, in, nor, of, on, or, the, to and up, which are usually
% not capitalized unless they are the first or last word of the title.
% Linebreaks \\ can be used within to get better formatting as desired.
% Do not put math or special symbols in the title.
\title{Robot-As-A-Sensor: Forming a Sensing Network with Robots for Underground Mining Missions}
%
%
% author names and IEEE memberships
% note positions of commas and nonbreaking spaces ( ~ ) LaTeX will not break
% a structure at a ~ so this keeps an author's name from being broken across
% two lines.
% use \thanks{} to gain access to the first footnote area
% a separate \thanks must be used for each paragraph as LaTeX2e's \thanks
% was not built to handle multiple paragraphs
%

\author{Xiaoyu~Ai,
        Chengpei~Xu,               
        Binghao~Li,~\IEEEmembership{Senior Member,~IEEE,}
        and~Feng~Xia,~\IEEEmembership{Senior Member,~IEEE}% <-this % stops a space
\thanks{Xiaoyu Ai, Chengpei Xu and Binghao Li are with the School of Minerals and Energy Resources Engineering, University of New South Wales, Sydney, Australia. Email addresses: \{x.ai, chengpei.xu, binghao.li\}@unsw.edu.au}% <-this % stops a space
\thanks{Feng Xia is with the School of Computing Technologies, Royal Melbourne Institute of Technology (RMIT) University, Melbourne, Australia.}% <-this % stops a space
\thanks{Correspondence: feng.xia@rmit.edu.au}% <-this % stops a space
%\thanks{Manuscript received XX XX, 2024; revised XX XX, 2024.}
}

% note the % following the last \IEEEmembership and also \thanks - 
% these prevent an unwanted space from occurring between the last author name
% and the end of the author line. i.e., if you had this:
% 
% \author{....lastname \thanks{...} \thanks{...} }
%                     ^------------^------------^----Do not want these spaces!
%
% a space would be appended to the last name and could cause every name on that
% line to be shifted left slightly. This is one of those "LaTeX things". For
% instance, "\textbf{A} \textbf{B}" will typeset as "A B" not "AB". To get
% "AB" then you have to do: "\textbf{A}\textbf{B}"
% \thanks is no different in this regard, so shield the last } of each \thanks
% that ends a line with a % and do not let a space in before the next \thanks.
% Spaces after \IEEEmembership other than the last one are OK (and needed) as
% you are supposed to have spaces between the names. For what it is worth,
% this is a minor point as most people would not even notice if the said evil
% space somehow managed to creep in.

% The paper headers
\markboth{Submitted to IEEE Transactions on Cognitive and Developmental Systems}%
{Ai \MakeLowercase{\textit{et al.}}: Robot-As-A-Sensor: Forming a Sensing Network with Robots for Underground Mining Missions}
% The only time the second header will appear is for the odd numbered pages
% after the title page when using the twoside option.
% 
% *** Note that you probably will NOT want to include the author's ***
% *** name in the headers of peer review papers.                   ***
% You can use \ifCLASSOPTIONpeerreview for conditional compilation here if
% you desire.

% If you want to put a publisher's ID mark on the page you can do it like
% this:
%\IEEEpubid{0000--0000/00\$00.00~\copyright~2015 IEEE}
% Remember, if you use this you must call \IEEEpubidadjcol in the second
% column for its text to clear the IEEEpubid mark.

% use for special paper notices
%\IEEEspecialpapernotice{(Invited Paper)}

% make the title area
\maketitle

% As a general rule, do not put math, special symbols or citations
% in the abstract or keywords.
\begin{abstract}
Nowadays, robots are deployed as mobile platforms equipped with sensing, communication and computing capabilities, especially in the mining industry, where they perform tasks in hazardous and repetitive environments. Despite their potential, individual robots face significant limitations when completing complex tasks that require the collaboration of multiple robots. This collaboration requires a robust wireless network to ensure operational efficiency and reliability. This paper introduces the concept of ``Robot-As-A-Sensor'' (RAAS), which treats the robots as mobile sensors within structures similar to Wireless Sensor Networks (WSNs). We later identify specific challenges in integrating RAAS technology and propose technological advancements to address these challenges. Finally, we provide an outlook about the technologies that can contribute to realising RAAS, suggesting that this approach could catalyse a shift towards safer, more intelligent, and sustainable industry practices. We believe that this innovative RAAS framework could significantly transform industries requiring advanced technological integration.
\end{abstract}

% Note that keywords are not normally used for peerreview papers.

% For peer review papers, you can put extra information on the cover
% page as needed:
% \ifCLASSOPTIONpeerreview
% \begin{center} \bfseries EDICS Category: 3-BBND \end{center}
% \fi
%
% For peerreview papers, this IEEEtran command inserts a page break and
% creates the second title. It will be ignored for other modes.
\IEEEpeerreviewmaketitle

\section{Introduction}
% The very first letter is a 2 line initial drop letter followed
% by the rest of the first word in caps.
% 
% form to use if the first word consists of a single letter:
% \IEEEPARstart{A}{demo} file is ....
% 
% form to use if you need the single drop letter followed by
% normal text (unknown if ever used by the IEEE):
% \IEEEPARstart{A}{}demo file is ....
% 
% Some journals put the first two words in caps:
% \IEEEPARstart{T}{his demo} file is ....
% 
% Here we have the typical use of a "T" for an initial drop letter
% and "HIS" in caps to complete the first word.
\IEEEPARstart{T}{he} fifth industrial revolution, as known as Industry 5.0, represents the current trends of advanced manufacturing technologies\cite{leng2022industry}. The pathway towards Industry 5.0 requires advanced networking, collaborative work of robots, intelligent systems, and a greater emphasis on sustainability and resilience\cite{xu2021industry}. This revolution aims to increase the efficiency and productivity of the overall manufacturing process and foster a more automated, safe and sustainable working environment for the workers and the surroundings\cite{huang2022industry}. The mining industry, serving as a critical supplier of raw materials like metals and coal, plays a pivotal role in manufacturing, energy production, construction and many other industry sectors. Therefore, upgrading the mining industry to meet the requirements of Industry 5.0 will be a significant shift in how mining operations are conducted, leveraging digital and autonomous technologies to enhance efficiency, safety, and environmental sustainability --- eventually entering the era of Mining 5.0\cite{chen2023mining}.

Robots equipped with multiple sensors can play a critical role in the evolution towards Mining 5.0 by obtaining vision-based data from cameras or Light Detection And Ranging (LiDAR) under normal and hazardous environments\cite{xu2024seeing}. These vision-based data can be further processed to extract information on the particular interests of the end users, such as texts, features and specific objects\cite{zhang2021rethinking,saydam2023feature,xu2024seeing}. Alternatively, post-sensing algorithms can be applied to serve the same purpose by extracting the semantic information from the acquired multimedia data\cite{xu2019lecture2note,xu2022semantic,xu2022morphtext}. These robots also utilise a range of non-vision-based sensors, such as infrared, ultrasonic, Inertial Measurement Units (IMUs) and wireless signal properties, to autonomously navigate and conduct operations in underground or open-pit mines and enhance safety and operational decision-making\cite{wang2019indoor,xu2022arbitrary,gu2022strokepeo,zhang2024v2vssc}. 
They are integral in automating complex tasks like drilling, excavation, and hauling, thereby reducing human involvement and improving productivity and safety. The integration of various sensors allows these robots to collect comprehensive data, supporting better decision-making and increasing the efficiency of mining operations.

Within the framework of a wireless network, a robot equipped with multiple sensors exemplifies a mobile, integrated sensing unit. These robots collect data and are tasked with disseminating this information to adjacent units of a similar nature or transmitting it back to the backhaul network for more in-depth analysis. This is precisely the role fulfilled by Wireless Sensor Networks (WSNs). WSNs are essentially networks of spatially distributed autonomous sensors that are tasked with monitoring various physical or environmental conditions, such as temperature, sound, and pressure. These sensors work collaboratively to route the collected data back to a central location through the network.

Initially conceived for military applications, particularly for battlefield surveillance, the use of WSNs has significantly broadened to encompass a diverse range of industrial and civilian domains. These include industrial process monitoring and control, which ensures the efficiency and safety of manufacturing operations; machine health monitoring, which helps in the early detection of equipment failures; environmental and habitat monitoring, which aids in the conservation of ecosystems; healthcare applications, which enhance patient care through continuous monitoring; home automation, which increases home security and energy efficiency; and traffic control, which improves vehicle flow and reduces congestion\cite{gulati2022review}. 

A local network is essential if robots are to perform increasingly automated and complex tasks collaboratively in hostile environments. This becomes particularly crucial for missions in underground mines. In such settings, a local network allows robots equipped with diverse sensors to communicate and share data seamlessly, regardless of the availability of high-bandwidth network infrastructure. This connectivity is vital for coordinating tasks, sharing sensory inputs, and making real-time collective decisions. However, it is important to note that there are significant differences between a conventional WSN and the desired network in this context:

\paragraph{Scalability} The desired network needs scalable technologies like Long-Range (LoRa) and Bluetooth to incorporate physical layer communications with various data rates complemented by algorithms for real-time network optimisation and predictive maintenance.
\paragraph{Mobility} The desired network should be capable of managing the roaming within the network with a minimised overhead to accommodate the random movements of the individual robots and reliance on single-path routing.
\paragraph{Communication Coverage} The desired network should achieve effective communication coverage in mines and requires strategically placed gateways and mesh networking due to the limited backhaul access to the surface.
\paragraph{Communication Latency} The desired network should also manage the latency of different types of acquired data, such as multimedia and telegraphic data. 

The challenges outlined above underscore the critical necessity for pioneering approaches to network design and management, especially in the context of deploying robots within the intricate and unpredictable settings of underground mines. Addressing these challenges demands innovative strategies that can adapt to the unique demands posed by such environments. In our forthcoming exploration, we aim to propose the concept of a network named ``Robot-As-A-Sensor'' (RAAS). The envisioned RAAS network is engineered to facilitate seamless collaboration among these sensor-laden robots, primarily relying on localised communications. Our investigation will delve into the intricacies of crafting such a network infrastructure to meet the stringent demands of underground mining operations. Furthermore, we will embark on an extensive examination to identify and evaluate the array of technologies that hold promise in actualising the envisioned RAAS network. By scrutinising the landscape of available technologies, we seek to ascertain the most viable solutions capable of furnishing the requisite functionality and robustness demanded by the RAAS network. Below, we provide a structured outline of this paper.
\begin{itemize}
    \item Section \ref{sec:review} will review the current progress of critical technologies that will be adopted in this RAAS network and demonstrate their technological benefits of using these technologies.
    \item Section \ref{sec: raas} will formally present the concept of the RAAS network and how this network will contribute to Mining 5.0.
    \item Sections \ref{sec:challenges} and \ref{sec:outlook} will analyse the challenges of realising the RAAS network and provide technical outlooks on the technologies that can contribute to realising a RAAS network. 
\end{itemize}

\section{Review of Current Progress}
\label{sec:review}
\subsection{Review of Sensing Technologies}
Sensing stands as one of the foundational technologies underpinning the efficacy of the RAAS network. Within this network, the sensing units seamlessly integrated into the robots serve as the linchpin for both communication and local processing functionalities. The versatility of these sensing units manifests in their adaptability to diverse task requirements, thereby endowing the RAAS network with the flexibility needed to complete the assigned tasks during underground mining operations. These sensing units can be broadly classified into two categories: vision-based and non-vision-based sensing. Vision-based sensors leverage sophisticated imaging technologies to capture and interpret visual data, thereby enabling robots to discern their surroundings via optical signals. Conversely, non-vision-based sensing includes but not limited to acoustic, thermal, proximity, and chemical sensors. By harnessing these disparate sensing modalities, robots within the RAAS network can gather a comprehensive spectrum of environmental data, facilitating robust decision-making processes and enhancing overall operational efficiency. Consequently, the selection and integration of sensing units tailored to the specific demands of underground mining tasks are pivotal in realizing the full potential of the RAAS network.

Vision-based sensing uses cameras to capture multimedia data for the next processing stage. The cameras can even be equipped with special filters to collect optical data at different wavelengths, from infrared and visible light to ultraviolet. For example, in \cite{zhou2015guidance}, a vision-based sensing system named ``Guidance'' was introduced. This system incorporates multiple stereo-sensing units to support basic visual tasks vital for robotic operation. It features built-in functions like a visual odometer for tracking the robot's movement, obstacle avoidance, and depth perception. This functionality is crucial for tasks that require precise navigation and environmental interaction. Vision-based sensing is also widely used in the mining industry. In \cite{szrek2020inspection}, robots equipped with infrared cameras are used to inspect belt conveyors in underground mines. These robots can detect overheated idlers and other potential mechanical failures, ensuring timely maintenance and reducing downtime. Such applications increase the efficiency of mining operations and enhance worker safety by minimising the need for human presence in dangerous areas. Besides using cameras to obtain visual data, LiDAR-based sensing technologies are also increasingly being utilised in the mining industry for various applications, from mineral exploration to mine closure and environmental monitoring. In \cite{trybala2022calibration}, multi-sensor wheeled robots are calibrated to inspect and create detailed 3D maps of underground mining tunnels, utilizing an array of sensors, including cameras and LiDARs. These maps are crucial for planning and risk management, allowing for better preparedness against potential mining hazards.

Non-vision-based sensing is supplementary to vision-based sensing with its additional features. In\cite{qadri2021automatic}, the development and implementation of an innovative automatic robotic scanning and inspection system specifically tailored for detecting carbon monoxide levels was reported, which aimed to enhance worker safety. As a valuable sensing technology, non-vision-based sensors can be a supplement to vision-based sensing in the context of multi-modal sensing. In\cite{yu2022all}, a multi-modal robotic sensing system featuring an all-printed soft electronic skin-based human-machine interface was presented. This advanced platform is designed for robots to make autonomous decisions. The technology can detect various hazardous materials, including explosives. Machine learning algorithms were applied to decode signals and remotely control the robot, enhancing safety and functionality in extreme or contaminated environments. 

In the view of communication networks, the key difference between vision-based and non-vision-based sensing is the demanded communication bandwidth to transmit and receive the obtained data. Data obtained from vision-based sensing is generally large (typically larger than several kilobytes) compared to non-vision-based sensing (typically several bytes). Simply using high-capacity communication technologies such as 5G cellular networks, which are the best solution, is only viable on the surface but not underground (Using cellular networks in underground mines is hindered by poor signal penetration due to the dense surrounding materials and rocks that weaken and even block radio frequency signals\cite{mishra2021}. Moreover, the complex and varied mine layouts require customised network designs far different from standard above-ground cellular networks, making standard deployment difficult and requiring innovative solutions tailored for the unique underground conditions\cite{moridi2018}). Therefore, the communication network should effectively accommodate the communication requirements of both sensing technologies. In the next subsection, we will introduce the state-of-the-art WSNs and how they can contribute to RAAS networks.

\subsection{Review of Wireless Sensing Networks}
% You must have at least 2 lines in the paragraph with the drop letter
% (should never be an issue)

As another fundamental technology in RAAS, WSNs are specifically designed to communicate sequential telegraphic data generated by sensors and offer several advantages for industrial applications and RAAS over Wi-Fi and cellular networks. WSNs are notably flexible and scalable, allowing for deployment in various environments. They can be expanded with minimal disruption and at minimal additional cost, facilitating the scaling of operations as needed\cite{kandris2020applications}. Besides, WSNs are designed to be energy-efficient, consuming minimal power\cite{mohamed2018survey} for both transmission and receiving. This makes them especially suitable for long-term monitoring tasks where sensors may rely on battery power or renewable energy sources, thus contributing to energy conservation. In realistic applications, WSNs are generally combined with the concept of meshed networks due to the large amount devices included in their operation area. A Meshed WSN is especially well-suited for sensor deployments that require flexibility and scalability. The interconnected devices can automatically establish and maintain mesh connectivity among themselves, thus providing multiple paths for data transmission to enhance the network's reliability. In~\cite{li2009sasa}, meshed WSNs facilitate real-time monitoring of environmental conditions and structural integrity within mines. Systems like Structure-Aware Self-Adaptive WSN systems utilise mesh networks to detect changes in mine structures, potentially preventing collapses by providing early warnings. In~\cite{shibalabala2020performance}, effective communication systems were tested to ensure functionality under the strain of additional nodes and environmental challenges typical to underground settings. In~\cite{bandyopadhyay2009wireless}, a wireless information and safety system using ZigBee-compliant active radio frequency identification (RFID) devices to form an IEEE 802.15.4-based mesh network was reported. This network is designed to locate, trace, and manage mobile assets and people and monitor different environmental conditions using sensors, providing crucial data for enhancing safety and operational efficiency in mines.

In the application of meshed WSNs underground, it is noted that devices can only be deployed alongside tunnels, and the transmission of radio signals is largely limited within tunnel spaces. Therefore, the sensor data are concentrated on some particular devices and propagate through the tunnel like a linearly connected network. In~\cite{branch2020lora}, a network utilising LoRa technology to establish WSNs arranged in a linear topology that is largely aligned with the tunnels, rather than the more common star topology of LoRaWAN, was reported. The linear network topology is particularly useful in scenarios where end-to-end communication needs to be relayed from outside the range of backhaul access, such as underground mining, remote monitoring of pipelines, or along borders and transportation routes. Key challenges in deploying LoRa-based linear networks include managing interference, ensuring reliable message forwarding across multiple relays, and optimising energy consumption and network lifespan. Various studies have explored these aspects, proposing solutions like gossip routing for efficient message propagation and energy-saving techniques to enhance the adaptability and robustness of these networks.

\section{Robot-As-A-Sensor}
\label{sec: raas}
The utilisation of robots equipped with an array of sensors has witnessed a notable surge within the mining industry, catalysing advancements in routine operations and handling hazardous environments. Yet, as the complexity of tasks escalates, a pressing need arises to fortify the existing infrastructure with a robust and efficient communication network. This imperative stems from the exigency to foster seamless data exchange among individual robots, facilitating real-time collaborative decision-making processes. Moreover, the dynamism inherent to mining environments necessitates the implementation of adaptive mechanisms wherein instructions issued to each robot can autonomously re-calibrate in response to shifting conditions. Consequently, conceptualising a dedicated Wireless Sensor Network (WSN) tailored specifically for these sensor-laden robots emerges as a pragmatic approach from a network engineering standpoint. This concept, the ``Robot-As-A-Sensor'' network, treats the robots as autonomous entities and integral components endowed with multifaceted sensing capabilities. By recognising the robots as active participants within the sensor network ecosystem, the RAAS paradigm transcends conventional notions of robot deployment, ushering in a paradigm shift characterised by enhanced coordination and adaptability. The RAAS network embodies a symbiotic fusion of robotics and wireless sensor technologies, emblematic of a paradigmatic leap towards realising autonomous and intelligent mining operations. 

\begin{figure*}[ht!]
    \centering
    \includegraphics[width=\textwidth]{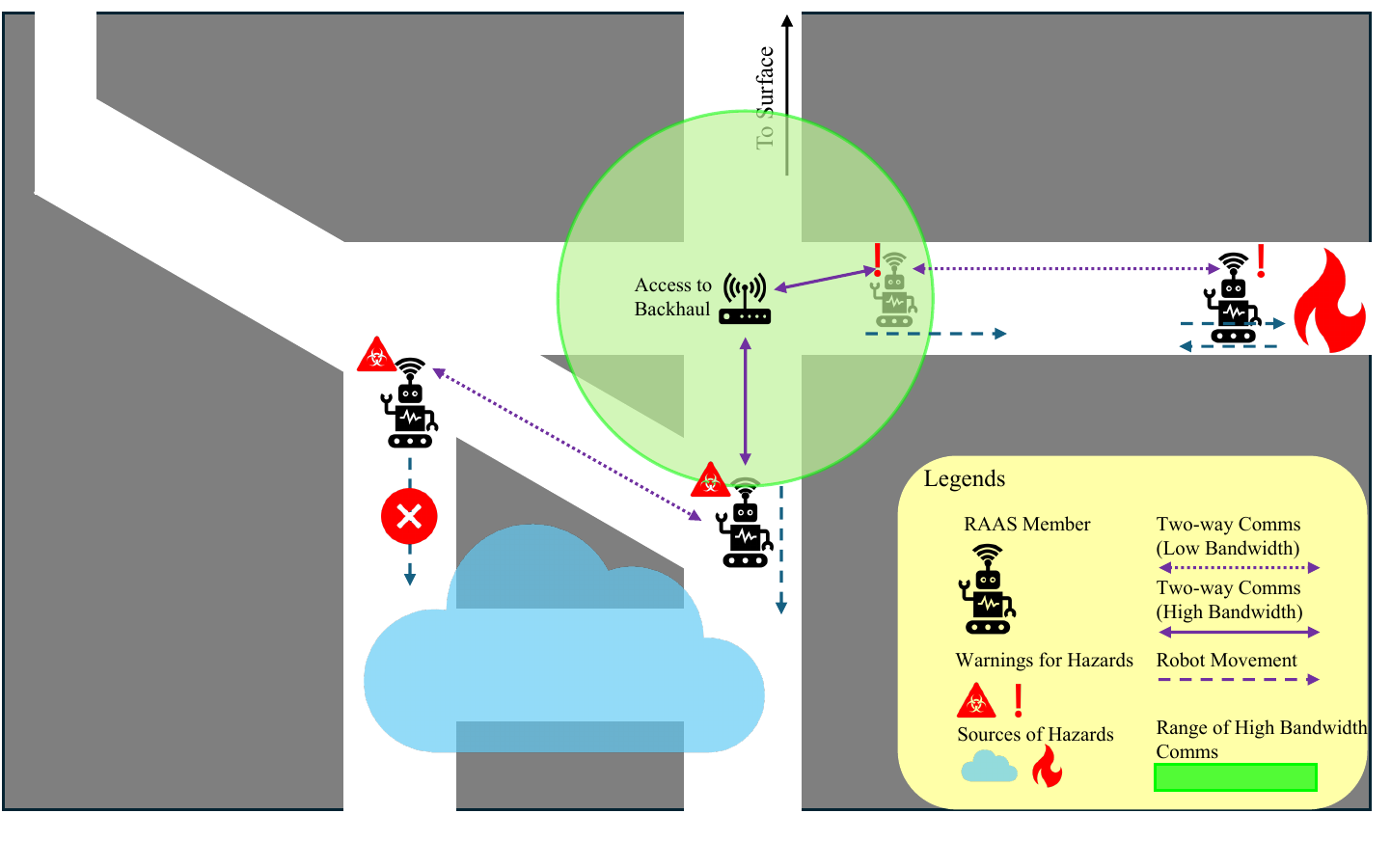}
    \caption{A graphical illustration of Robot-As-A-Sensor for underground mine sites. The dashed lines indicate the paths and patterns that the robots follow during inspections. In a dynamic environment, robots can adapt these paths based on real-time data collected from other robots or their own sensors in the operation area. The solid arrows represent the high bandwidth communication channels established between the robots, which can be used for transmitting multimedia data. The dotted arrows represent low-bandwidth communications that might be used for telegraphing sensor data and/or control signals. The green shaded area depicts the area where the backhaul network is accessible. Warnings for Hazards and Sources of Hazards Critical for automated inspections in potentially hazardous environments. Robots can detect and communicate hazards to each other, enhancing situational awareness and safety.}
    \label{fig:raas_diagram}
\end{figure*}

In Fig.~\ref{fig:raas_diagram}, portable RAAS members are deployed within the challenging environment of an underground mine site. Their the movements are along predefined paths but the paths can be adapted based on real-time sensory data from other robots or their own sensors. A crucial aspect highlighted in Fig.~\ref{fig:raas_diagram} is the two-way communication capability of the RAAS network. It distinguishes between high-bandwidth communication channels, used for transmitting multimedia data, and low-bandwidth channels, typically utilised for sensor data or control signals. This communication infrastructure allows for seamless data exchange between robots and with the broader backhaul network. Fig.~\ref{fig:raas_diagram} also demonstrates the importance of detecting and communicating hazards within the mine environment. Robots can identify potential hazards based on their sensor data and share this information with each other, enhancing overall situational awareness and operational safety. Overall, the RAAS network demonstrates the versatility and resilience of robotic networks in dynamic and hazardous environments, which will improve both safety and operational efficiency in underground mining operations. 

Readers may find that the illustration in Fig.~\ref{fig:raas_diagram} shares a similar illustration of a typical WSN. However, we note that there are three major differences between a RAAS network and a typical WSN: 

1). Most WSNs assume that the end devices within a WSN have less mobility - the devices are either at a fixed location or move slowly around the sinking device node (A sinking node in a WSN refers to a device node that collects data transmitted from various sensor nodes distributed throughout the network. This node usually aggregates and/or processes the data before delivering it to the backhaul network (if it has access to the backhaul network)\cite{liu2012performance}. The placement of the sink node will influence overall network performance and sensor data availability). However, this assumption generally does not apply to a RAAS network since each RAAS member may transmit and receive data as long as any two RAAS members are within their own communication ranges and do not rely on the sinking device nodes to concentrate the collected data, which makes the RAAS network highly flexible and robust against unexpected communication outages in some parts of the entire network.

2). A typical WSN relies on a single type of radio signals for the actual data communication due to power saving or coverage requirements, but RAAS may take advantage of multiple radio signals for different communication purposes. This hybrid communication in the RAAS network will facilitate a layered approach to data exchange, allowing the system to adapt dynamically. Each RAAS member can prioritise and switch between various radio signals (even frequency channels) based on specific Quality of Service (QoS) requirements or the depth of information required.

3).  WSNs typically handle telegraphic data, which consists of small-sized, regularly transmitted data packets, often not exceeding a few kilobytes. This nature of data is less demanding in terms of bandwidth, allowing for efficient transmission over limited bandwidth channels. In contrast, RAAS networks manage a more complex data ecosystem. RAAS encompasses a hybrid mix of data types, including audio sequences, multimedia data, and telegraphic data. This diversity in data types should be accommodated with the unique QoS requirements associated with each type. Each member within an RAAS network must consider these requirements and tailor their data transmission strategies accordingly. Besides, optimising these protocols involves adjusting to the bandwidth needs and considering other QoS parameters such as jitter, error rates, and reliability. 

Realising RAAS networks promises to elevate the robustness, efficiency of collaborative automation and real-time data sharing among robots equipped with diverse sensory capabilities. This shift optimises operational workflows and significantly mitigates human exposure to hazardous environments. As the mining industry continues to evolve, the strategic deployment of RAAS networks will be crucial in harnessing the full spectrum of benefits offered by advanced robotic sensing and communication technologies. The next section will analyse the critical challenges of realising a RAAS network for underground missions.

\section{Challenges of Realising RAAS in Underground Mine Sites}
\label{sec:challenges}
\subsection{Scalability}
Enhancing the dynamic adaptability of IoT networks, which include LoRa, Narrow Band-IoT (NB-IoT), and Bluetooth technologies, requires specialised strategies that cater to the specific properties of each communication protocol. In the context of LoRa, dynamic power management and adaptive data rate functions are critical to this adaptability. Dynamic power management helps achieve an efficient balance between energy use and dependable communication by modulating the transmission power according to the distance between nodes. Simultaneously, adaptive data rate features automatically adjust transmission speeds to suit various communication conditions effectively. Additionally, real-time network optimisation algorithms and predictive maintenance techniques are vital, continuously monitoring signal integrity, detecting interference, and anticipating maintenance requirements, thereby enhancing the overall adaptability of LoRa networks.

The protocol design and communication bandwidth can limit the scalability of the overall network. A scalable RAAS network often involves heterogeneous devices with varying capabilities and network conditions and is expected to adapt its bit rate to accommodate these variations, which becomes challenging as the number of devices increases. In a RAAS network, each robot, equipped with multiple sensors, generates significant data that needs to be communicated to other nodes for processing and coordination. If the communication bit rate is not sufficiently high, it becomes a bottleneck, restricting the amount of data that can be transmitted at any given time and thereby limiting the network's ability to scale effectively. For a (slotted) time-division communication system, The overall throughput $T$ for a network with $n$ members can be approximated by\cite{namislo1984analysis},
\begin{equation}
\label{eq:throughput}
    T = \frac{p_t p_s E[s]}{E[t]}\,,
\end{equation}
where $p_t$ is the probability that at least one node transmits in a given slot time, $p_s$ is the probability of a successful transmission, given that there is at least one transmission, $E[s]$ is the average packet payload size and is the average time the medium is occupied as a result of transmission. The variables $p_t$ and $p_s$ can be expressed as\cite{namislo1984analysis},
\begin{eqnarray}
     p_t =& 1 - (1 - \tau)^n \,,\label{eq:pt}\\
     p_s =& \frac{n \tau (1 - \tau)^{n-1}}{p_t}\,,\label{eq:ps}
\end{eqnarray}
where $\tau$ is the probability that a specific node decides to transmit in a given slot. Recalling Eqs.~\ref{eq:pt} and~\ref{eq:ps}, it is clear that the throughput $T$ will decrease when $n$ becomes large due to the collision of each packets. Consequently, if the network does not have a coordinated mechanism to manage multiple access (like CSMA/CD used in Ethernet or CSMA/CA used in Wi-Fi), the likelihood of data collision increases. When a collision occurs, the transmitted data packets are typically corrupted or lost and must be resent. 

The impact of limited bit rate on scalability is further exacerbated when the network includes a mix of devices with different data transmission requirements. For example, some robots might be equipped with high-definition cameras that generate large volumes of multimedia data, while others might only need to transmit smaller amounts of sensor data. Without a sufficiently high bit rate, the network might struggle to accommodate the high data demands of the multimedia-rich devices while also efficiently servicing the lower data needs of other devices.

%For NB-IoT networks, key features like dynamic resource allocation and advanced mobility management protocols are essential for adapting to variable environments. Dynamic resource allocation optimises bandwidth use by tailoring resources to the prevailing demand, while enhanced mobility management ensures uninterrupted communication as devices move through diverse settings. Furthermore, the integration of firmware over-the-air updates allows NB-IoT devices to update and adapt to new demands autonomously continuously. 

\subsection{Mobility}
Taking the mobility of network members into consideration of network design is important for resource management, data routing, localisation, context-aware services, and network planning. This proactive approach enhances the reliability, responsiveness, and overall performance of the RAAS network across diverse applications and environments, enabling stakeholders to unlock the full potential of the network for improved user experiences and operational efficiency.

Forming a robust wireless network for end devices with random movement has always been a hot research topic in networking and wireless communications, particularly for vehicle-to-vehicle (V2V) communications\cite{camp2002survey,alkhatieb2020performance}. The Ad hoc On-Demand Distance Vector (AODV) protocol is a popular routing protocol studied among various protocols in the literature\cite{perkins2003ad,chakeres2004aodv}. The AODV protocol is a reactive routing protocol used primarily in Mobile Ad hoc NETworks (MANETs). Unlike proactive protocols that maintain routing information for every node in the network at all times, AODV creates routes only when necessary. Features of this protocol include\cite{chakeres2004aodv}:
\paragraph{Route Recovery} When a node needs to send a packet but lacks an existing route to the destination, it initiates by broadcasting a Route Request (RREQ) to its immediate neighbors. This RREQ contains the destination and source IP addresses, a unique broadcast ID, and the destination's last known sequence number to ensure route freshness. Upon receiving the RREQ, Neighbouring nodes check if they've previously processed this request, if they have a route to the destination, or if they are the destination itself. If none of these apply, they increment the hop count on the RREQ and forward it. Should the RREQ reach the destination or a node with a recent route to the destination, a Route Reply (RREP) is generated and sent back to the source using the reverse path established by the RREQ, during which each node updates its routing table with the new route information.

\paragraph{Route Maintenance} Once a route is established, it is maintained until a link failure is detected. If a node detects a break in the link to the next hop on the route to the destination, it generates a Route Error (RERR) message. This RERR is sent to all upstream nodes reliant on the broken link, prompting them to update their routing tables to remove any paths through the failed link. If these nodes still require the route, they may initiate a new route discovery process to establish an alternative path.

The AODV protocol (and its variants) may offer advantages when RAAS are deployed underground, primarily through its on-demand route discovery and topology self-maintenance nature\cite{temene2022survey}. However, directly implementing AODV in the context of RAAS will lead to several major challenges. The standard AODV protocol typically introduces a larger overhead than typical WSN protocols. In power-constrained scenarios such as RAAS, larger overhead leads to unnecessary power consumption and eventually limits the operation time of the individuals within the network. Besides, due to its route discovery mechanism, the AODV protocol can face significant performance issues in high-data applications, which increases control traffic and network congestion. Thus, while AODV is adaptable and beneficial in specific scenarios, its application in resource-limited environments like RAAS requires careful consideration and modifications to optimise the overall network performance.
 
\subsection{Communication Coverage}
Although a RAAS network is expected to operate independently when deployed underground, the strategic placement of the access to backhaul is critical for achieving real-time data collection. Generally, the maximum communication distance is subject to the channel loss that is commonly given by\cite{tse2005fundamentals},
\begin{equation}
\label{eq:path_loss}
    L(d) = L_0 + 10 \gamma \log_{10} \left(\frac{d}{d_0}\right) + X_{\sigma}\,,
\end{equation}
where $L(d)$ is the path loss (in dB) as a function of the distance, $d$, between a transmitter and a receiver, $ L_0$ is the path loss at the reference distance, $d_0$ (typically 1 meter), $\gamma$ is the path loss exponent, and $X_{\sigma}$ is a Gaussian random variable representing shadow fading, with a mean of zero and a standard deviation of $\sigma$. With Eq.~\ref{eq:path_loss}, the maximum communication distance, $d_m$, can be obtained by solving the equation for $d_m$,
\begin{eqnarray}
\label{eq:max_dist}
    P^{min}_r &= P_t - L_0 - 10 \gamma \log_{10} \left(\frac{d_m}{d_0}\right) - X_{\sigma}\,,
\end{eqnarray}
where $P^{min}_r$ is the minimum receiving power of the receiver, $P_t$ is the transmission power. The path loss can differ for different mine sites and radio frequency signals. In\cite{branch2022measurements}, The LoRa radio propagation at 915 MHz in an underground gold mine demonstrates that LoRa technology provides excellent propagation both with and without line of sight, benefiting from waveguide effects inside the tunnels, which causes less attenuation to LoRa signal transmission than free-space.  

From Eq.~\ref{eq:max_dist}, it seems straightforward that $d_m$ can be increased by increasing the transmission power $P_t$. However, it should be noted that simply increasing the transmission power to enhance communication distances does not necessarily improve overall network performance. This is because increasing transmission power can lead to a higher level of interference devices closer to the transmitter, especially in dense networks. This higher interference can reduce the network's capacity by affecting the quality of other ongoing communications. The relationship between the transmission power, interference, and the channel capacity can be obtained by\cite{lovasz1979shannon,tse2005fundamentals},
\begin{equation}
    C=B\log_2\cdot (1+S)\,,
\end{equation}
where $C$ is the channel capacity in bits per second (bps), $B$ is the bandwidth of the signal in Hertz (Hz) and $S$ is the Signal-to-Interference-Noise Ratio (SINR)\cite{tse2005fundamentals,mogensen2007lte},
\begin{equation}
    S = \frac{P^\prime_t}{P_n+P_I}\,,
\end{equation}
where $P^\prime_t$ is the transmission power other than the transmission power of the interfering device, $P_n$ is the noise power, and $P_I=P_t$ is the interference power. For spread-spectrum-based signals such as LoRa, a strong interference will reduce the SINR at any receivers within the network, leading to erroneous packets and re-transmission.

The purpose of extending the network coverage is twofold. First, employing mesh networking techniques effectively bypasses limitations in direct sinking node connectivity, allowing nodes to pass messages between each other and thereby extending their operational range. This approach ensures that even if only a subset of devices has backhaul access, the network can still transmit data to the back-end at the expense of additional delays. This concept is exemplified in underground mining operations in\cite{branch2020lora}. Second, extending the network coverage with higher bandwidth will benefit the overall performance more. High-bandwidth wireless networks such as Wi-Fi can support over 100 Mbps\footnote{The IEEE 802.11ac standard is assumed.} but only within a range of approximately 25 metres\cite{ASLAM2024173}. Extending the Wi-Fi coverage to cover the whole underground site would be ideal, but it would be subject to the infrastructural nature of the mine sites. Conventional WSNs can be used to extend communication coverage at a lower bit rate. For example, LoRa offers a robust physical layer framework for establishing a LoRa WSN. Similar connectivity issues are evident in other WSN technologies like NB-IoT.

\subsection{Communication Latency}
Minimising the communication latency is a important but difficult challenge when realising RAAS. As discussed in the previous subsection, extending the communication coverage is usually at the price of adding additional latency to the data transmission. WSNs are typically categorised as low-bandwidth networks, compared to the state-of-the-art mobile and Wi-Fi networks. LoRa, operating in sub-gigahertz bands, limits its data rates to between 0.3 and 27 kbps\cite{noreen2017study}. Bluetooth, evolving through various versions for improved speed and range, allows for Basic Rate/Enhanced Data Rate up to 3 Mbps\cite{afonso2016performance}. However, we note that the above WSN technologies are tailored for specific use cases: LoRa for low-data-rate, long-distance applications and Bluetooth for short-range, high-data-rate consumer electronics.

However, in the context of RAAS, robots carry sensors generating telegraphic data and multimedia data obtained via cameras. Transferring the multimedia data via conventional WSNs will induce a large delay (For example, sending a JPEG image of 1MB will lead to a latency of 4 seconds from one Bluetooth device to another. The 4-second latency is obtained by assuming the maximum bit rate (2 Mbps) of Bluetooth communications without channel impairments). Transmitting multimedia data over low-bandwidth networks involves complex challenges such as interference leading to quality degradation, which can be mitigated through effective management strategies like controlling transmission power and re-transmissions~\cite{murayama2007}. Efficient source coding/decoding algorithms can be useful to reduce the bitrate requirement for multimedia data\cite{civanlar1997}. Real-time routing protocols can be used to address congestion and burst traffic, with techniques such as adaptive traffic shaping and multi-path forwarding enhancing Quality of Service\cite{ahmed2017}. Furthermore, multimedia applications, particularly video streaming, produce variable bit-rate streams that conflict with the constant-bit-rate channels of real-time protocols, requiring integrated control of compression parameters and network bandwidth\cite{silvestreblanes2011}.

\section{Technical Outlook of RAAS Network Design}
\label{sec:outlook}
\begin{figure*}[ht!]
    \centering
    \includegraphics[width=\textwidth]{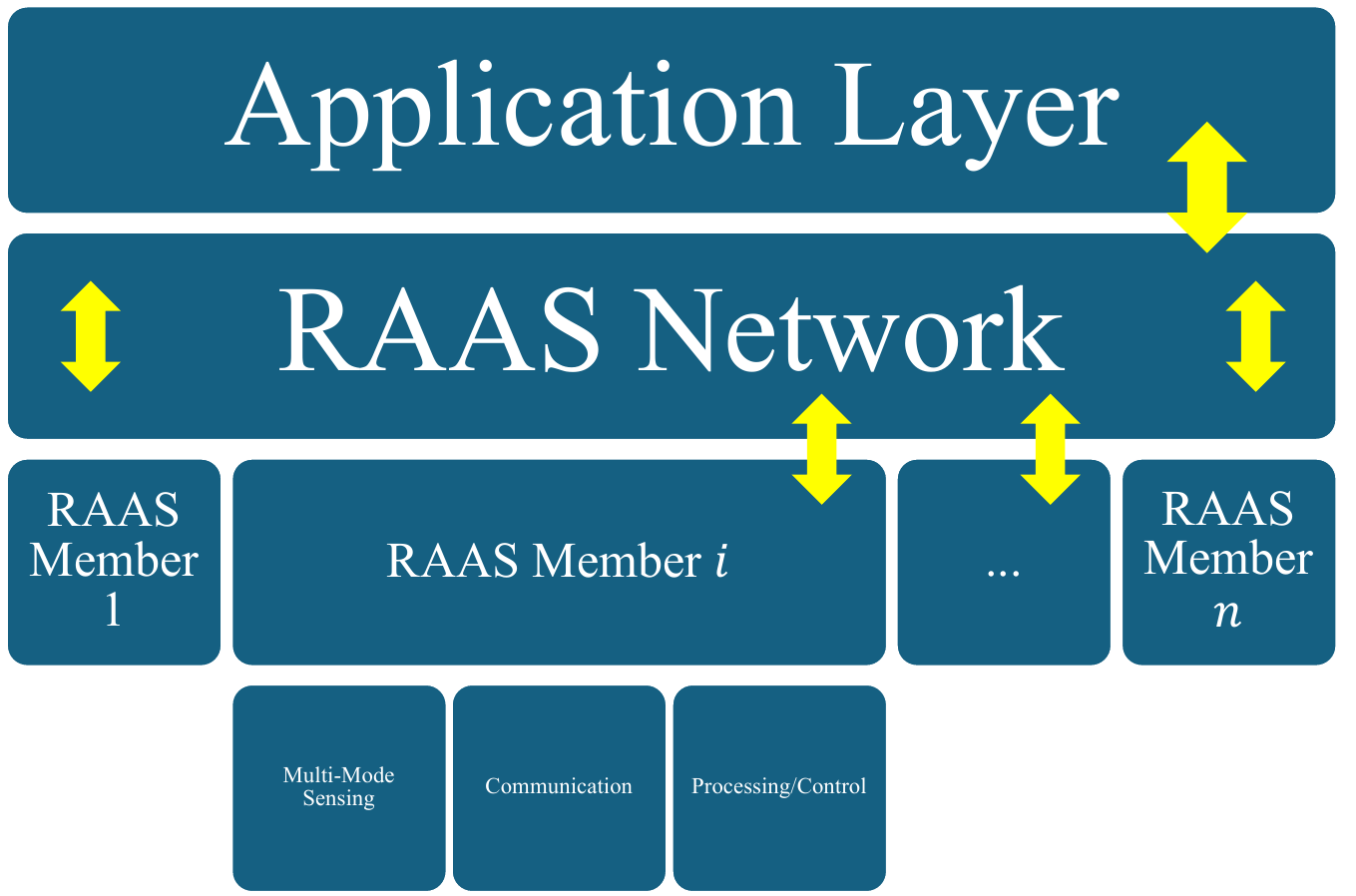}
    \caption{An illustration of one of the possible architectures of RAAS Networks. The illustration shows that an RAAS network effectively integrates multiple RAAS members to perform specific tasks and share data crucial for operational success in challenging underground environments. For each RAAS member within the network (e.g., RAAS Member $i$), it is capable of conducting multi-mode sensing and communications between neighbouring members and dynamically controlling the progress of existing missions, which is particularly designed to operate within the complex and hazardous confines of an underground mine.}
    \label{fig:raas_architecture}
\end{figure*}

Based on the discussion in the last section, it is clear that the central to the functionality of these robots is the communication capability, but the sensing and local data process capabilities are also important to the RAAS network. Fig.~\ref{fig:raas_architecture} demonstrates a possible architecture of RAAS network.

\subsection{Integrated Sensing and Communication}
Integrated Sensing and Communication (ISAC) is proposed to be a core component of 5G-Advanced and 6G networks\cite{liu2021}. ISAC proposes efficient spectrum use and cost-effective hardware solutions to integrate communication and sensing capabilities. As the ISAC technology advances, it introduces the potential for these networks to perform dual roles, greatly enhancing the functionality and efficiency of mobile communication systems.

The advantages of integrating ISAC into the RAAS network is threefold. First, ISAC transforms IoT devices by enabling them to ``perceive and connect"\cite{liu2021}. This capability is crucial for implementing RAAS networks as it allows for seamless data sharing and decision-making. Distributed RAAS members, through networked collaboration, improve environmental sensing by decentralised transmission and processing. The fusion of data from multiple sources mitigates sensing inaccuracies and boosts the efficiency of distributed sensing. Consequently, sensing tasks can be distributed effectively among nodes, and integrating sensing with communication facilitates a collective understanding of the environment. 

Second, ISAC is pivotal in meeting the varied demands of future network scenarios that will predominantly include a range of devices such as sensors and robots\cite{ouyang2022performance}. These devices depend on precise sensing and positioning data to make informed decisions and manage operations intelligently. ISAC overcomes the limitations of traditional wireless sensor networks by merging sensing and communication into a unified function, thus tightly linking the physical and digital dimensions of network nodes. This integration addresses the challenges of traditional networks that typically separate communication and sensing processes.

Third, traditional IoT networks often follow a ``communication-after-sensing'' model which can lead to delays in data exchange and increased processing costs\cite{cui2021integrating}. In contrast, ISAC-enabled devices are capable of ``sensing-with-communication,'' which reduces both time delays and processing costs. This makes ISAC particularly effective in large-scale networks surpassing traditional sensor networks' capabilities by integrating the physical and digital aspects of IoT nodes more closely.

Significant progress has been made in the development of ISAC, with the integration of sensing capabilities into cellular networks becoming a definitive feature of 6G\cite{ouyang2022performance}. This allows the networks to utilize dense cellular infrastructure to build a perceptive network environment, improving both communication and sensing capabilities. Technological advancements focus on optimizing ISAC signals and processing methods to address the inherent trade-offs between sensing accuracy and communication effectiveness\cite{wei2023}.

However, challenges such as interference management and signal optimisation remain. Current research efforts are directed towards overcoming these hurdles to refine ISAC performance further, ensuring it meets the stringent demands of future wireless networks\cite{lu2023}. These efforts are critical in moving towards fully integrated and efficient systems that support the sophisticated applications anticipated in the 6G era\cite{wild2021}.

%ISAC systems offer several enhancements over traditional WSNs by combining sensing and communication functionalities into a single platform, making ISAC better suitible for forming the RAAS network. Such integration not only simplifies the architecture but also optimizes the use of spectrum and hardware resources, leading to cost reductions and improved performance. ISAC facilitates dual functionality, where communication assists sensing and vice versa, enhancing overall network efficacy\cite{liu2021}. Additionally, it promotes resource efficiency and energy savings, particularly beneficial in applications requiring both high-performance sensing and communication, such as autonomous vehicles and smart cities\cite{Cui2021,Pottie2000}. This makes ISAC a key enabler for future wireless systems with enhanced capabilities and efficiencies.

\begin{table*}[!t]
\centering
\caption{Interconnection between the challenges and possible technical solutions.}
\label{table:challenges_solutions}
\begin{tabular}{>{\raggedright\arraybackslash}m{3cm}|>{\raggedright\arraybackslash}m{14cm}}
\hline
\hline
\textbf{Challenges} & \textbf{Technical Solutions} \\ 
\hline
Scalability & 
- Integrated Sensing and Communication (ISAC) enhances the functionality and efficiency of mobile communication systems, allowing dense cellular infrastructure to build a perceptive network environment. \\

\hline
Mobility &
- Advanced network protocols accommodate the movement of RAAS members.

- EC (EC) reduces latency and improves decision-making speed by processing data at the RAAS members. \\
\hline
Communication Coverage &
- Semantic Communications optimises bandwidth usage by focusing on the semantic meaning of transmitted data, enhancing the relevance and efficiency of communications. \\
\hline
Communication Latency &
- ISAC is used for its dual functionality, optimising spectrum and hardware resources and improving overall network efficacy.

- EC local processing reduces transmission delays and conserves bandwidth by minimising data transfer.

- Semantic Communications ensure that only information with significant semantic value is communicated, optimizing bandwidth and improving communication efficiency. \\
\hline
\hline
\end{tabular}
\end{table*}

\subsection{Edge Computing and Semantic Communications}
In the context of RAAS networks, Edge Computing (EC) can reduce communications costs by processing the collected data before its transmission\cite{Qiu2020}. Data that needs to be transmitted over long distances may be reduced due to local processing. Additionally, edge computing optimises resource utilisation and manages network traffic efficiently, preventing congestion\cite{Zhou2021Energy-Efficient}. Techniques like optimised offloading also contribute to bandwidth savings by balancing data processing needs with network capabilities, ensuring efficient data transmission and enhancing overall network performance.

In the context of RAAS networks, adopting EC has been instrumental in enhancing the reliability and security of mine data operations and eventually speeds up data processing and fortifies privacy and security measures, which are critical in mining operations\cite{Jian2021}. Additionally, EC facilitates resource optimisation in intelligent coal mining by addressing the interaction demands of smart devices in underground settings, thereby improving overall efficiency and revenue generation\cite{Cheng2023}. Another groundbreaking application of EC in mining involves predictive analysis of geological data. Leveraging deep learning models, such as Long Short-Term Memory (LSTM) networks, combined with EC enhances the performance of mineral exploration by optimising the analysis of drilling data. This leads to more accurate predictions and efficient resource exploration, which are vital for the mining industry\cite{Yin2020}.

Semantic communication is a breakthrough beyond traditional communication paradigms, focusing on transmitting the meaning (or semantics) of messages rather than the raw bit strings. This approach enhances the efficiency and relevance of communication by ensuring that the receiver understands the intended meaning, leveraging techniques like artificial intelligence for semantic extraction and encoding. Semantic communication evaluates success based on the accuracy of the meaning transmitted rather than traditional error rates used in conventional systems \cite{qin2021}. In the context of the RAAS network, semantic communications can be useful when combined with EC. EC can support semantic communication by facilitating faster semantic encoding and decoding through local processing, which is essential for maintaining the semantic integrity of the messages in dynamic environments\cite{yu2023}. This approach prioritises context and semantic fidelity, allowing redundant information to be omitted and reduce the data payload. Techniques such as semantic transformations introduce acceptable data loss at the bit level but preserve the integrity of the meaning, thus conserving bandwidth. Additionally, integrating EC for local semantic processing minimises the volume of data transmitted over the network, further enhancing bandwidth efficiency. Overall, semantic communications optimise bandwidth use by ensuring that only information with significant semantic value is communicated, mirroring the efficiency of human communication, which prioritises meaning over detailed syntactic accuracy.

\section{Conclusion}
In this work, we presented the concept of Robot-As-A-Sensor for advancing the efficiency and safety of underground mining operations through a sophisticated network of robotic sensors. We established the unique demands of deploying RAAS systems to tackle challenges such as limited mobility, communication coverage, and latency issues inherent to subterranean environments. By integrating robotic systems with WSNs, we demonstrated the potential to enhance operational safety and efficiency, reduce risks, and significantly decrease the need for human exposure in dangerous mining zones. Furthermore, exploring potential solutions, including ISAC and EC, highlights our forward-looking approach to embracing future technological advancements. These technologies pave the way for improved data processing speeds and reduced latency, which are crucial for real-time decision-making in critical mining operations. As we conclude, it is imperative to recognise that developing RAAS systems requires multidisciplinary inputs and innovations. Ongoing research is encouraged to substantially improve how sensory data is collected, processed, and utilised, thereby fostering safer and more efficient underground mining practices. The future of RAAS in underground mining evolves with the advent of emerging technologies that offer significant enhancements in communication and sensory accuracy. Continuous improvements in this field will likely provide invaluable tools for mining operators, potentially setting new standards in mining operations and safety.

% if have a single appendix:
%\appendix[Proof of the Zonklar Equations]
% or
%\appendix  % for no appendix heading
% do not use \section anymore after \appendix, only \section*
% is possibly needed

% use appendices with more than one appendix
% then use \section to start each appendix
% you must declare a \section before using any
% \subsection or using \label (\appendices by itself
% starts a section numbered zero.)
%

% use section* for acknowledgment
\section*{Acknowledgment}
This work is supported by the Cooperative Research Centres Projects (CRC-P) Grant Project named "Underground mine LoRa network for monitoring/control/backup/rescue/robotics".

% Can use something like this to put references on a page
% by themselves when using endfloat and the captionsoff option.
\ifCLASSOPTIONcaptionsoff
  \newpage
\fi

% trigger a \newpage just before the given reference
% number - used to balance the columns on the last page
% adjust value as needed - may need to be readjusted if
% the document is modified later
%\IEEEtriggeratref{8}
% The "triggered" command can be changed if desired:
%\IEEEtriggercmd{\enlargethispage{-5in}}

% references section

% can use a bibliography generated by BibTeX as a .bbl file
% BibTeX documentation can be easily obtained at:
% http://mirror.ctan.org/biblio/bibtex/contrib/doc/
% The IEEEtran BibTeX style support page is at:
% http://www.michaelshell.org/tex/ieeetran/bibtex/
%\bibliographystyle{IEEEtran}
\bibliographystyle{mybst}
% argument is your BibTeX string definitions and bibliography database(s)
\bibliography{IEEEabrv, references}

\begin{thebibliography}{10}

\bibitem{leng2022industry}
J.~Leng, W.~Sha, B.~Wang, P.~Zheng, C.~Zhuang, Q.~Liu, T.~Wuest, D.~Mourtzis, and L.~Wang, ``Industry 5.0: Prospect and retrospect,'' {\em Journal of Manufacturing Systems}, vol.~65, 279--295, 2022.

\bibitem{xu2021industry}
X.~Xu, Y.~Lu, B.~Vogel-Heuser, and L.~Wang, ``Industry 4.0 and industry 5.0—inception, conception and perception,'' {\em Journal of manufacturing systems}, vol.~61, 530--535, 2021.

\bibitem{huang2022industry}
S.~Huang, B.~Wang, X.~Li, P.~Zheng, D.~Mourtzis, and L.~Wang, ``Industry 5.0 and society 5.0—comparison, complementation and co-evolution,'' {\em Journal of manufacturing systems}, vol.~64, 424--428, 2022.

\bibitem{chen2023mining}
L.~Chen, J.~Xie, X.~Zhang, J.~Deng, S.~Ge, and F.-Y. Wang, ``Mining 5.0: Concept and framework for intelligent mining systems in cpss,'' {\em IEEE Transactions on Intelligent Vehicles}, 2023.

\bibitem{xu2024seeing}
C.~Xu, H.~Fu, L.~Ma, W.~Jia, C.~Zhang, F.~Xia, X.~Ai, B.~Li, and W.~Zhang, ``Seeing text in the dark: Algorithm and benchmark,'' {\em arXiv preprint arXiv:2404.08965}, 2024.

\bibitem{zhang2021rethinking}
Y.-f. Zhang, J.~Zheng, L.~Li, N.~Liu, W.~Jia, X.~Fan, C.~Xu, and X.~He, ``Rethinking feature aggregation for deep rgb-d salient object detection,'' {\em Neurocomputing}, vol.~423, 463--473, 2021.

\bibitem{saydam2023feature}
S.~Saydam, C.~Xu, B.~Li, B.~Topal, and S.~Saydam, ``Feature sampling and balancing for detecting rock bolts from the lidar point clouds,'' in {\em ISRM Congress}, ISRM--15CONGRESS, ISRM, 2023.

\bibitem{xu2019lecture2note}
C.~Xu, R.~Wang, S.~Lin, X.~Luo, B.~Zhao, L.~Shao, and M.~Hu, ``Lecture2note: Automatic generation of lecture notes from slide-based educational videos,'' in {\em 2019 IEEE International Conference on Multimedia and Expo (ICME)}, 898--903, IEEE, 2019.

\bibitem{xu2022semantic}
C.~Xu, W.~Jia, R.~Wang, X.~He, B.~Zhao, and Y.~Zhang, ``Semantic navigation of powerpoint-based lecture video for autonote generation,'' {\em IEEE Transactions on Learning Technologies}, vol.~16, no.~1, 1--17, 2022.

\bibitem{xu2022morphtext}
C.~Xu, W.~Jia, R.~Wang, X.~Luo, and X.~He, ``Morphtext: Deep morphology regularized accurate arbitrary-shape scene text detection,'' {\em IEEE Trans. Multimedia}, 2022.

\bibitem{wang2019indoor}
Z.~Wang, P.~Sokliep, C.~Xu, J.~Huang, L.~Lu, and Z.~Shi, ``Indoor position algorithm based on the fusion of wifi and image,'' in {\em 2019 Eleventh International Conference on Advanced Computational Intelligence (ICACI)}, 212--216, IEEE, 2019.

\bibitem{xu2022arbitrary}
C.~Xu, W.~Jia, T.~Cui, R.~Wang, Y.-f. Zhang, and X.~He, ``Arbitrary-shape scene text detection via visual-relational rectification and contour approximation,'' {\em IEEE Trans. Multimedia}, 2022.

\bibitem{gu2022strokepeo}
Z.~Gu, X.~Yang, W.~Jia, C.~Xu, P.~Yu, X.~He, H.~Chen, and Y.~Lin, ``Strokepeo: Construction of a clinical ontology for physical examination of stroke,'' in {\em 2022 9th International Conference on Digital Home (ICDH)}, 218--223, IEEE, 2022.

\bibitem{zhang2024v2vssc}
Y.~Zhang, J.~Li, K.~Luo, Y.~Yang, J.~Han, N.~Liu, D.~Qin, P.~Han, and C.~Xu, ``V2vssc: A 3d semantic scene completion benchmark for perception with vehicle to vehicle communication,'' {\em arXiv preprint arXiv:2402.04671}, 2024.

\bibitem{gulati2022review}
K.~Gulati, R.~S.~K. Boddu, D.~Kapila, S.~L. Bangare, N.~Chandnani, and G.~Saravanan, ``A review paper on wireless sensor network techniques in internet of things (iot),'' {\em Materials Today: Proceedings}, vol.~51, 161--165, 2022.

\bibitem{zhou2015guidance}
G.~Zhou, L.~Fang, K.~Tang, H.~Zhang, K.~Wang, and K.~Yang, ``Guidance: A visual sensing platform for robotic applications,'' in {\em Proceedings of the IEEE Conference on Computer Vision and Pattern Recognition Workshops}, 9--14, 2015.

\bibitem{szrek2020inspection}
J.~Szrek, J.~Wodecki, R.~Blazej, and R.~Zimroz, ``An inspection robot for belt conveyor maintenance in underground mine—infrared thermography for overheated idlers detection,'' {\em Applied Sciences}, vol.~10, no.~14, 4984, 2020.

\bibitem{trybala2022calibration}
P.~Trybała, J.~Szrek, F.~Remondino, J.~Wodecki, and R.~Zimroz, ``Calibration of a multi-sensor wheeled robot for the 3d mapping of underground mining tunnels,'' {\em The International Archives of the Photogrammetry, Remote Sensing and Spatial Information Sciences}, 2022.

\bibitem{qadri2021automatic}
I.~Qadri, A.~Muneer, and S.~M. Fati, ``Automatic robotic scanning and inspection mechanism for mines using iot.,'' in {\em IOP conference series: Materials science and engineering}, vol.~1045, 012001, IOP Publishing, 2021.

\bibitem{yu2022all}
Y.~Yu, J.~Li, S.~A. Solomon, J.~Min, J.~Tu, W.~Guo, C.~Xu, Y.~Song, and W.~Gao, ``All-printed soft human-machine interface for robotic physicochemical sensing,'' {\em Science robotics}, vol.~7, no.~67, eabn0495, 2022.

\bibitem{mishra2021}
P.~Mishra, A.~Swain, S.~Kumar, and S.~K. Mandal, ``Wireless paging system for underground mines,'' {\em Radioelectronics and Communications Systems}, vol.~64, 14--25, 2021.

\bibitem{moridi2018}
M.~A. Moridi, M.~Sharifzadeh, Y.~Kawamura, and H.~Jang, ``Development of wireless sensor networks for underground communication and monitoring systems (the cases of underground mine environments),'' {\em Tunnelling and Underground Space Technology}, vol.~73, 127--138, 2018.

\bibitem{kandris2020applications}
D.~Kandris, C.~Nakas, D.~Vomvas, and G.~Koulouras, ``Applications of wireless sensor networks: an up-to-date survey,'' {\em Applied system innovation}, vol.~3, no.~1, 14, 2020.

\bibitem{mohamed2018survey}
R.~E. Mohamed, A.~I. Saleh, M.~Abdelrazzak, and A.~S. Samra, ``Survey on wireless sensor network applications and energy efficient routing protocols,'' {\em Wireless Personal Communications}, vol.~101, 1019--1055, 2018.

\bibitem{li2009sasa}
M.~Li and Y.~Liu, ``Underground coal mine monitoring with wireless sensor networks,'' in {\em ACM Trans. Sens. Networks}, vol.~5, 1--29, 2009.

\bibitem{shibalabala2020performance}
J.~Shibalabala and T.~Swart, ``Performance analysis of wireless mesh networks for underground mines,'' in {\em 2020 International Conference on Artificial Intelligence, Big Data, Computing and Data Communication Systems (icABCD)}, 1--6, 2020.

\bibitem{bandyopadhyay2009wireless}
L.~Bandyopadhyay, S.~K. Chaulya, P.~Mishra, A.~Choure, and B.~M. Baveja, ``Wireless information and safety system for mines,'' {\em Safety Science}, vol.~47, no.~5, 616--624, 2009.

\bibitem{branch2020lora}
P.~Branch, B.~Li, and K.~Zhao, ``A lora-based linear sensor network for location data in underground mining,'' in {\em Telecom}, vol.~1, 6, MDPI, 2020.

\bibitem{liu2012performance}
W.~Liu, K.~Lu, J.~Wang, G.~Xing, and L.~Huang, ``Performance analysis of wireless sensor networks with mobile sinks,'' {\em IEEE transactions on vehicular technology}, vol.~61, no.~6, 2777--2788, 2012.

\bibitem{namislo1984analysis}
C.~Namislo, ``Analysis of mobile radio slotted aloha networks,'' {\em IEEE Journal on Selected Areas in Communications}, vol.~2, no.~4, 583--588, 1984.

\bibitem{camp2002survey}
T.~Camp, J.~Boleng, and V.~Davies, ``A survey of mobility models for ad hoc network research,'' {\em Wireless communications and mobile computing}, vol.~2, no.~5, 483--502, 2002.

\bibitem{alkhatieb2020performance}
A.~AlKhatieb, E.~Felemban, and A.~Naseer, ``Performance evaluation of ad-hoc routing protocols in (fanets),'' in {\em 2020 IEEE wireless communications and networking conference workshops (WCNCW)}, 1--6, IEEE, 2020.

\bibitem{perkins2003ad}
C.~Perkins, E.~Belding-Royer, and S.~Das, ``Ad hoc on-demand distance vector (aodv) routing,'' tech. rep., 2003.

\bibitem{chakeres2004aodv}
I.~D. Chakeres and E.~M. Belding-Royer, ``Aodv routing protocol implementation design,'' in {\em 24th International Conference on Distributed Computing Systems Workshops, 2004. Proceedings.}, 698--703, IEEE, 2004.

\bibitem{temene2022survey}
N.~Temene, C.~Sergiou, C.~Georgiou, and V.~Vassiliou, ``A survey on mobility in wireless sensor networks,'' {\em Ad Hoc Networks}, vol.~125, 102726, 2022.

\bibitem{tse2005fundamentals}
D.~Tse and P.~Viswanath, {\em Fundamentals of wireless communication}.
\newblock Cambridge university press, 2005.

\bibitem{branch2022measurements}
P.~Branch, ``Measurements and models of 915 mhz lora radio propagation in an underground gold mine,'' {\em Sensors}, vol.~22, no.~22, 2022.

\bibitem{lovasz1979shannon}
L.~Lov{\'a}sz, ``On the shannon capacity of a graph,'' {\em IEEE Transactions on Information theory}, vol.~25, no.~1, 1--7, 1979.

\bibitem{mogensen2007lte}
P.~Mogensen, W.~Na, I.~Z. Kov{\'a}cs, F.~Frederiksen, A.~Pokhariyal, K.~I. Pedersen, T.~Kolding, K.~Hugl, and M.~Kuusela, ``Lte capacity compared to the shannon bound,'' in {\em 2007 IEEE 65th vehicular technology conference-VTC2007-Spring}, 1234--1238, IEEE, 2007.

\bibitem{ASLAM2024173}
M.~Aslam, X.~Jiao, W.~Liu, M.~Mehari, T.~Havinga, and I.~Moerman, ``A novel hardware efficient design for ieee 802.11ax compliant ofdma transceiver,'' {\em Computer Communications}, vol.~219, 173--181, 2024.

\bibitem{noreen2017study}
U.~Noreen, A.~Bounceur, and L.~Clavier, ``A study of lora low power and wide area network technology,'' in {\em 2017 International Conference on Advanced Technologies for Signal and Image Processing (ATSIP)}, 1--6, IEEE, 2017.

\bibitem{afonso2016performance}
J.~A. Afonso, A.~J.~F. Maio, and R.~Simoes, ``Performance evaluation of bluetooth low energy for high data rate body area networks,'' {\em Wireless Personal Communications}, vol.~90, 121--141, 2016.

\bibitem{murayama2007}
S.~Murayama and F.~Tobagi, ``Multimedia data transmission over wireless network with interference,'' {\em IEICE Trans. Commun.}, vol.~90-B, 651--659, 2007.

\bibitem{civanlar1997}
M.~Civanlar and A.~Reibman, ``Signal processing for networked multimedia,'' {\em IEEE Signal Process. Mag.}, vol.~14, 39--41, 1997.

\bibitem{ahmed2017}
A.~A. Ahmed, ``A real-time routing protocol with adaptive traffic shaping for multimedia streaming over next-generation of wireless multimedia sensor networks,'' {\em Pervasive Mob. Comput.}, vol.~40, 495--511, 2017.

\bibitem{silvestreblanes2011}
J.~Silvestre-Blanes, L.~Almeida, R.~Marau, and P.~Pedreiras, ``Online qos management for multimedia real-time transmission in industrial networks,'' {\em IEEE Transactions on Industrial Electronics}, vol.~58, 1061--1071, 2011.

\bibitem{liu2021}
F.~Liu, Y.~Cui, C.~Masouros, J.~Xu, T.~Han, Y.~C. Eldar, and S.~Buzzi, ``Integrated sensing and communications: Toward dual-functional wireless networks for 6g and beyond,'' {\em IEEE Journal on Selected Areas in Communications}, vol.~40, 1728--1767, 2021.

\bibitem{ouyang2022performance}
C.~Ouyang, Y.~Liu, and H.~Yang, ``Performance of downlink and uplink integrated sensing and communications (isac) systems,'' {\em IEEE Wireless Communications Letters}, vol.~11, no.~9, 1850--1854, 2022.

\bibitem{cui2021integrating}
Y.~Cui, F.~Liu, X.~Jing, and J.~Mu, ``Integrating sensing and communications for ubiquitous iot: Applications, trends, and challenges,'' {\em IEEE Network}, vol.~35, no.~5, 158--167, 2021.

\bibitem{wei2023}
Z.~Wei, H.~Qu, Y.~Wang, X.~Yuan, H.~Wu, Y.~Du, K.~Han, N.~Zhang, and Z.~Feng, ``Integrated sensing and communication signals toward 5g-a and 6g: A survey,'' {\em IEEE Internet of Things Journal}, vol.~10, 11068--11092, 2023.

\bibitem{lu2023}
S.-J. Lu, F.~Liu, Y.~Li, K.~Zhang, H.~Huang, J.~Zou, X.~Li, Y.~Dong, F.~Dong, J.~Zhu, Y.~Xiong, W.~Yuan, Y.~Cui, and L.~Hanzo, ``Integrated sensing and communications: Recent advances and ten open challenges,'' {\em ArXiv}, vol.~abs/2305.00179, 2023.

\bibitem{wild2021}
T.~Wild, V.~Braun, and H.~Viswanathan, ``Joint design of communication and sensing for beyond 5g and 6g systems,'' {\em IEEE Access}, vol.~9, 30845--30857, 2021.

\bibitem{Qiu2020}
T.~Qiu, J.~Chi, X.~Zhou, Z.~Ning, M.~Atiquzzaman, and D.~O. Wu, ``Edge computing in industrial internet of things: Architecture, advances and challenges,'' {\em IEEE Communications Surveys \& Tutorials}, vol.~22, 2462--2488, 2020.

\bibitem{Zhou2021Energy-Efficient}
X.~Zhou, X.~he~Yang, J.~Ma, and K.~Wang, ``Energy-efficient smart routing based on link correlation mining for wireless edge computing in iot,'' {\em IEEE Internet of Things Journal}, vol.~9, 14988--14997, 2021.

\bibitem{Jian2021}
W.~Jian, Q.~Shi-jia, C.~Xiao-lin, and F.~Li-li, ``Research on reliability of mine data based on blockchain and edge computing,'' in {\em 2021 6th International Symposium on Computer and Information Processing Technology (ISCIPT)}, 399--404, 2021.

\bibitem{Cheng2023}
W.~Cheng, X.~Liu, and G.~Nie, ``Task offloading and resource allocation method for edge computing in intelligent coal mining,'' in {\em 2023 IEEE/CIC International Conference on Communications in China (ICCC Workshops)}, 1--6, 2023.

\bibitem{Yin2020}
J.~Yin, X.~Luo, Y.~Zhu, W.~Wang, L.~Wang, C.~Huang, and J.-H. Wang, ``An edge computing‐based predictive evaluation scheme toward geological drilling data using long short‐term memory network,'' {\em Transactions on Emerging Telecommunications Technologies}, vol.~32, 2020.

\bibitem{qin2021}
Z.~Qin, X.~Tao, J.~Lu, and G.~Y. Li, ``Semantic communications: Principles and challenges,'' {\em ArXiv}, 2021.

\bibitem{yu2023}
W.~li~Yu and J.~Zhao, ``Semantic communications, semantic edge computing, and semantic caching with applications to the metaverse and 6g mobile networks,'' {\em 2023 IEEE 43rd International Conference on Distributed Computing Systems (ICDCS)}, 2023.

\end{thebibliography}
%
% <OR> manually copy in the resultant .bbl file
% set second argument of \begin to the number of references
% (used to reserve space for the reference number labels box)

% biography section
% 
% If you have an EPS/PDF photo (graphicx package needed) extra braces are
% needed around the contents of the optional argument to biography to prevent
% the LaTeX parser from getting confused when it sees the complicated
% \includegraphics command within an optional argument. (You could create
% your own custom macro containing the \includegraphics command to make things
% simpler here.)

%\begin{IEEEbiography}[{\includegraphics[width=1in,height=1.25in,clip,keepaspectratio]{Bio/xiaoyu.JPG}}]{Xiaoyu Ai received his BEng. from Xidian University, China in 2013. He also received a MEngSc. and Ph.D. from the University of New South Wales, Australia, in 2015 and 2022, respectively. He is currently a post-doctoral researcher with the University of New South Wales, Australia. His research interests includes wireless communications, channel coding, wireless sensor networks and quantum communications.}
% or if you just want to reserve a space for a photo:

%\end{IEEEbiography}

% You can push biographies down or up by placing
% a \vfill before or after them. The appropriate
% use of \vfill depends on what kind of text is
% on the last page and whether or not the columns
% are being equalized.

%\vfill

% Can be used to pull up biographies so that the bottom of the last one
% is flush with the other column.
%\enlargethispage{-5in}

% that's all folks
\end{document}